\documentclass[aip,jap,reprint,superscriptaddress]{revtex4-1}
\usepackage{physics}
\usepackage{tikz, circuitikz}
\usepackage{graphicx}% Include figure files
\usepackage{dcolumn}% Align table columns on decimal point
\usepackage{bm} % bold math
\usepackage{hyperref} % add hypertext capabilities
\usepackage{cleveref} % better referencing
\usepackage{caption, subcaption} % Add subfigure environment
\usepackage{afterpage} % Figure positioning
\usepackage{xparse}

\newcommand{\fstop}{\text{.}} % Full stop in equation environment
\newcommand{\cma}{\text{,}} % Comma in equation environment
\begin{document}
	
	%\preprint{}
	
	\title{Circuit-theoretic phenomenological model of an electrostatic gate-controlled bi-SQUID}
	
	
	\author{Thomas X.\ Kong}
	% \email[Email: ]{thomas.kong@adelaide.edu.au}
	\author{Jace Cruddas}
	\author{Jonathan Marenkovic}
	\author{Wesley Tang}
	\affiliation{Department of Physics, The University of Adelaide, Adelaide, South Australia 5005, Australia}
	\affiliation{Quantum and Nano Technology Group (QuaNTeG), School of Chemical Engineering, The University of Adelaide, North Terrace Campus, Adelaide, 5000, South Australia, Australia}
	\author{Giorgio De Simoni}
	\affiliation{NEST, Istituto Nanoscienze-CNR and Scuola Normale Superiore, I-56127 Pisa, Italy}
	
	%\affiliation{Department of Physics, The University of Adelaide, Adelaide, South Australia 5005, Australia}
	%\affiliation{School of Chemical Engineering and Advanced Materials, The University of Adelaide, Adelaide, South Australia 5005, Australia}
	\author{Francesco Giazotto}
	\affiliation{NEST, Istituto Nanoscienze-CNR and Scuola Normale Superiore, I-56127 Pisa, Italy}
	\author{Giuseppe C.\ Tettamanzi}
	\email[Corresponding author, email: ]{giuseppe.tettamanzi@adelaide.edu.au}
	\affiliation{Quantum and Nano Technology Group (QuaNTeG), School of Chemical Engineering, The University of Adelaide, North Terrace Campus, Adelaide, 5000, South Australia, Australia}
	%\thanks{}
	
	
	\date{\today}
	
	\begin{abstract}
		A numerical model based on a lumped circuit element approximation for a bi-superconducting quantum interference device (bi-SQUID) operating in the presence of an external magnetic field is presented in this paper. Included in the model is the novel ability to capture the resultant behaviour of the device when a strong electric field is applied to its Josephson junctions by utilising gate electrodes. The model is used to simulate an all-metallic SNS (Al-Cu-Al) bi-SQUID, where good agreement is observed between the simulated results and the experimental data. The results discussed in this work suggest that the primary consequences of the superconducting field effect induced by the gating of the Josephson junctions are accounted for in our minimal model; namely, the suppression of the junctions super-current. Although based on a simplified semi-empirical model, our results may guide the search for a microscopic origin of this effect by providing a means to model the voltage response of gated SQUIDs. Also, the possible applications of this effect regarding the operation of SQUIDs as ultra-high precision sensors, where the performance of such devices can be improved via careful tuning of the applied gate voltages, are discussed at the end of the paper.
	\end{abstract}
	
	
	\maketitle
	
	\section{Introduction}
	Superconducting quantum interference devices (SQUIDs, a device with only two resistively shunted Josephson junctions connected in parallel via a loop of superconducting material)~\cite{Kornev2009, Mukhanov2014, Susan-PhD, Longhini2012, patent} have long been utilised as ultra-high precision magnetic flux-to-voltage transducers across a plethora of applications spanning medical imaging, remote sensing, geophysical surveying and quantum metrology~\cite{Tinkham, Susan-PhD, Jeanneret2009, Winkler2003}. While their detection capabilities greatly surpass that of classical sensors and magnetometers~\cite{Mukhanov2014,Lenz1990,Robbes2006}, their performance in terms of response linearity, i.e. their ability to not distort the electromagnetic signals they sense, is still far from being optimised. One of the main avenues for improving the performance of SQUID-based sensors is to connect a large number of SQUID cells in an array structure, a field under significant active investigation~\cite{Oppenlander2005, Oppenlander2012, Mitchell2016, Kornev2016, Kornev2009b, Mukhanov2014, Gio23}.\\
	
	%has the advantage of removing the need of large and potentially bulky arrays, or
	
	%to be engineered to exhibit highly linear voltage responses presented 
	
	An alternative approach for the design of optimal SQUIDs is to seek novel single-cell device geometries that exhibit an improved performance over that of a DC SQUID. This may provide an alternative to the DC SQUID as the repeating unit in complicated array structures, leading to further improvements to performance. Indeed, since its inception in 2009, the bi-SQUID design as a single-cell device has shown great potential as a novel and promising candidate for an optimised SQUID sensor exhibiting highly linear voltage responses~\cite{Kornev2009}. A bi-SQUID consists of three Josephson junctions connected within two loops of superconducting material. Two of the junctions are resistively shunted and are operated in the dissipative regime, which provide the driving contributions to the flux-dependent voltage response of the device. The third junction is typically un-shunted, and mainly serves to provide a nonlinear inductive contribution to the dynamics of the device. The quantum interference of the three junctions combined yield a voltage-flux response of greater linearity compared to that of a conventional DC SQUID.\\
	
	Initial theoretical studies demonstrated that even a small series array of bi-SQUIDs can achieve a response linearity of 120 dB~\cite{Kornev2009}. However, attempts to fabricate and operate such an ideal device using traditional tunnel-based Josephson junctions in practice have been largely unsuccessful in achieving the theorised response linearity~\cite{DeSimoni2022}. It is believed that the main limitations arise from the large inductance and capacitance of a typical tunnel junction~\cite{DeSimoni2022}. These issues can be circumvented by adopting a superconducting-normal metal-superconducting (SNS) junction architecture, which in addition have excellent ease-of-reproducibility in the fabrication process, allowing for precise control over the junction parameters~\cite{DeSimoni2019, DeSimoni2022}.	It was recently experimentally discovered that the application of a sufficiently strong electric field to an SNS junction results in the suppression of the junction critical current~\cite{DeSimoni2018}. Such observations could not be accounted for within the framework of the Bardeen-Cooper-Schrieffer (BCS) theory of superconductivity, and as a consequence, understanding the microscopic mechanism responsible for this unconventional superconducting field effect remains an open question in the discipline~\cite{DeSimoni2018, DeSimoni2021b, Rocci2020, DeSimoni2019, Paolucci2021, Sankar2022, Basset2021, Amoretti2022, haxell.etal_2023}.\\
	
	Regardless of the nature of the underlying mechanism~\cite{Basset21}, the superconducting field effect presents a means to precisely tune the critical current of the junctions in a superconducting device. This provides greater control over the device's behaviour, which can be exploited to yield an overall improved performance~\cite{DeSimoni2021b}.\\
	
	Numerical models for SQUIDs have been extensively utilised to assist in the design process of SQUIDs. They are used to computationally probe the parameter space for optimal device geometries and favourable operational regimes~\cite{Susan-PhD, Oppenlander2012, Chesca2004, Longhini2012}. However, owing to being a newly discovered phenomenon, the superconducting field effect is yet to be implemented within the framework of existing SQUID models. Therefore, this work proposes a minimal extension to existing SQUID models~\cite{Susan-PhD, Longhini2012, Kornev2009} that is able to capture the physics of gated Josephson junctions in the context of a SQUID. The approach introduced in this work is purely phenomenological; the application of a gate voltage to the junctions of a SQUID primarily results in the suppression of the junction critical current only. Also demonstrate that the inclusion of this effect is sufficient to describe the majority of effect observed in recently fabricated gated bi-SQUIDs. \\ %Indeed, it is very reassuring to mention that a good agreement between the theory and the experiments can be observed in~\Cref{fig:sim} and~\Cref{fig:main} of this document.  \\
	
	The SNS bi-SQUIDs to be considered and the experimental setup used to characterise their behaviour are introduced in~\Cref{sec:methodsA}. The main details regarding the model are described in~\Cref{sec:methodsB}. Simulation results of the devices are shown alongside corresponding experimental measurements in~\Cref{sec:results}. The results of the model are compared and validated versus the experiments in~\Cref{sec:discussion}. This section~\Cref{sec:discussion} concludes with some general discussion on how this effect may be exploited to improve device performance for sensing applications.
	
	\section{Methods}
	\label{sec:methods}
	\subsection{Device Fabrication and Measurement}
	\label{sec:methodsA}
	The bi-SQUIDs devices studied in this work are all-metallic, with a superconducting-normal-superconducting (SNS) type architecture for the Josephson junctions, and fabricated using single electron beam lithography (EBL) and a two-angle shadow-mask metal deposition technique~\cite{DeSimoni2022}. The bulk superconducting loops are constructed from Al, enclosing approximate areas of \( 22\pm 0.7\,\mu\text{m}^2 \) and \( 2\pm 0.7\,\mu\text{m}^2 \). The weak links are realised by proximitised mesoscopic nanowires constructed from Cu. \\
	
\begin{figure*}
		\begin{subfigure}{0.49\textwidth}
			\includegraphics[width=0.5\linewidth,keepaspectratio]{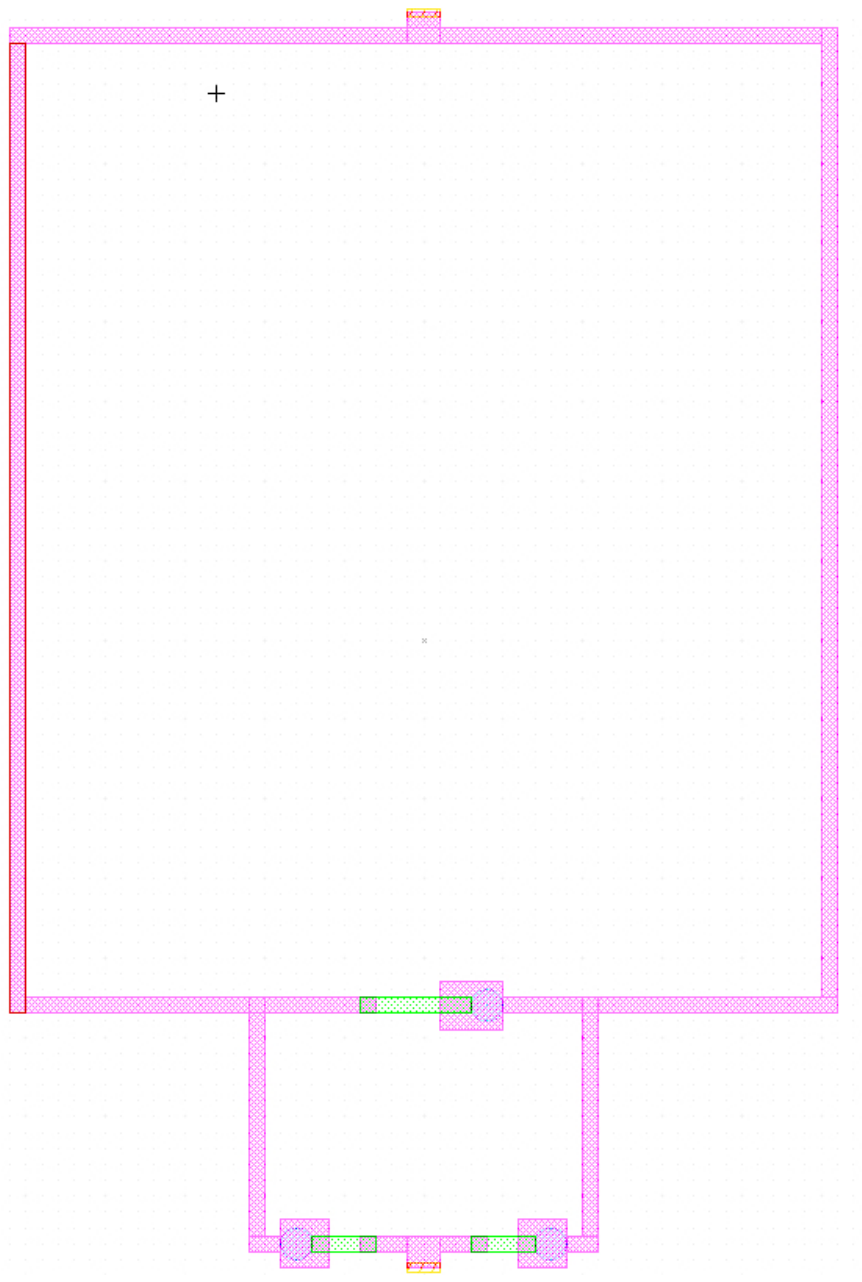}
			\caption{}
		\end{subfigure}\
		\begin{subfigure}{0.49\textwidth}
			\includegraphics[width=\linewidth,keepaspectratio]{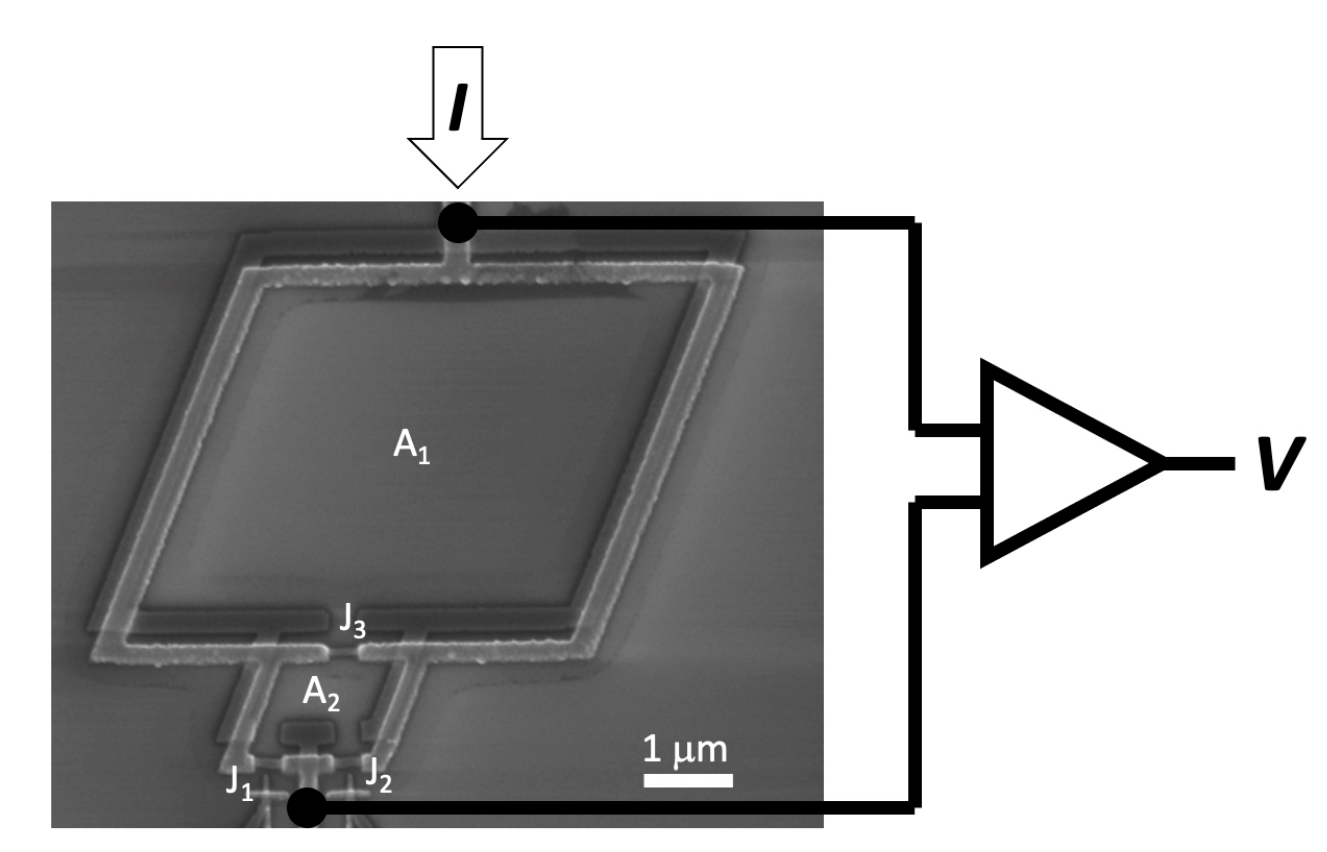}
			\caption{}
		\end{subfigure}
	
		\label{fig:gds}
	\caption{(a) Design file image of the bi-SQUID as viewed in the CAD software~\emph{Layout Editor}. (b) Scanning electron microscope (SEM) image of our all-metallic SNS bi-SQUID. The SNS junctions are labelled~\( J_1 \), \( J_2 \) and~\( J_3 \), with electrostatic gates placed in the vicinity of the~\( J_1 \) and~\( J_2 \) junctions, only. The ratios of the loop areas are such that \( \flatfrac{A_1}{A_2}\sim 10 \). An input current \( I \) is used to bias the device into the dissipative regime, which allows the measurement of a nonzero voltage response \( V \) across the bi-SQUID. }
	\label{fig:device}
	\end{figure*}

	The SNS bi-SQUIDs were fabricated by electron-beam lithography (EBL) and angle-resolved e-beam evaporation. The latter exploited a two-angle shadow-mask metal deposition through a suspended PMMA resist mask onto a Si/SiO2 substrate. The Al/Cu SN clean interfaces were obtained through electron-beam evaporation at a pressure $5\times10^{-11}$ Torr. The N section of the SNS junction consisted of a Ti/Cu bilayer evaporated perpendicular to the substrate, in which Ti and Cu have a thickness of 5 and 25 nm, respectively. The Ti is used as a sticking layer. The superconducting loop was then deposited, without breaking the vacuum, by evaporating a 100 nm-thick Al film at an angle of  13° with respect to the sample azimuthal axis. The lift-off procedure in acetone followed by sample rinsing in isopropanol completed the fabrication. A pair of Cu gate electrodes are positioned schematically below the bottom two SNS junctions of the device, labelled $J_{1}$ and $J_{2}$, as shown in the SEM image in~\Cref{fig:device}.\\ 
	
	The bi-SQUID used in this work consisting of SNS Josephson junctions boasts a number of generally advantageous properties such as a reproducible fabrication process and further tailoring and control of the current-phase relation. The primary advantage when compared to others in the literature, however, is that other than Dayem bridges which are known to have poor performance, SNS junctions are the only fully metallic Josephson junctions allowing for the control of the critical current via gate action; e.g.~\cite{Kornev2009}\\
	
	Cryogenic electrical characterization: The electrical measurement of the devices was carried out in a low-pass filtered $^3$He-$^4$He closed-cycle dilution fridge equipped with a superconducting electromagnet. Current-voltage (IV) measurements were carried out by current-biasing the devices by a room-temperature voltage generator in series with a 1 M$\Omega$ resistor. The 4-wire voltage drop across the interferometer was measured with a room-temperature pre-amplifier. Switching current values were retrieved by averaging the switching points of 15 repetitions of the same IV. The voltage vs. flux characterization was achieved through a low-frequency lock-in technique in a 4-wire configuration, where the ac bias current was set through the lock-in amplifier sinusoidal reference signal in series with a load resistor. \\
	
	Measurement of the characteristic behaviour of the device is carried out at cryogenic temperatures using a standard setup~\cite{DeSimoni2022, DeSimoni2021b}. A 17 Hz sinusodial input bias current signal of magnitude 23 \( \mu \)A RMS is applied to the device. External magnetic flux is applied to the device via a superconducting electromagnet. The lack of a microscopic model of the behaviour of a gated Josephson junction does not allow for predictive analysis of the modifications to noise within bi-SQUIDs as a result of gating. The experimental exploration of this effect has been left for future work.
	
	\afterpage{
	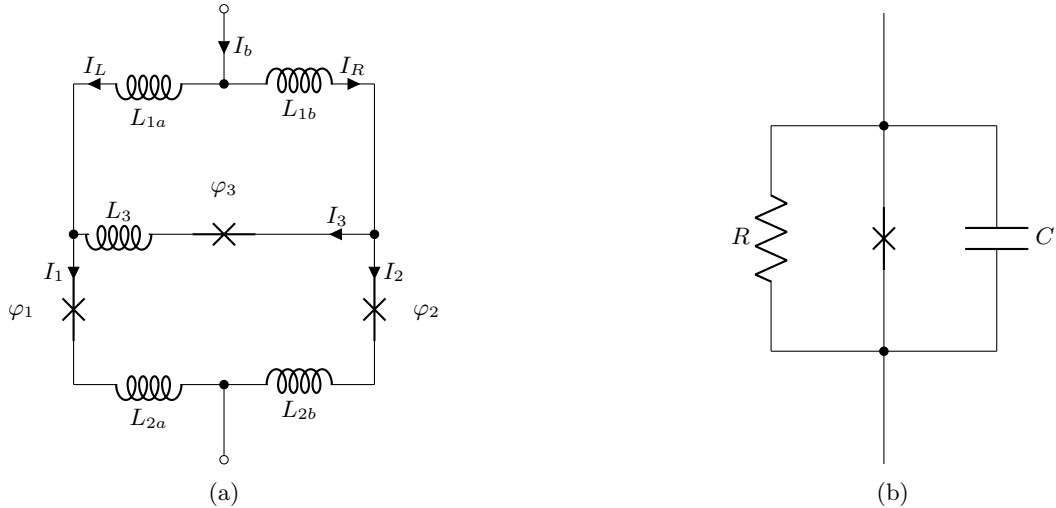
\begin{figure*}
		\begin{subfigure}{0.49\textwidth}
			\begin{circuitikz}
				% Input lead
				\draw (0,5) to[short,o-*, i=\( I_b \)] (0,4);
				
				% Top inductances
				\draw (0,4) to[L,i_=\( I_L \),l=\( L_{1a} \)] (-2,4);
				\draw (0,4) to[L,i^=\( I_R \),l_=\( L_{1b} \)] (2,4);
				\draw (2,2)
				to[short,i_=\( I_3 \)] (1,2)
				to[barrier,l_=\( \varphi_{3} \)] (-1,2)
				to[L,l_=\( L_{3} \)] (-1.8,2)
				to[short,-*] (-2,2);
				\draw (0,0) to[L,l=\( L_{2a} \)] (-2,0);
				\draw (0,0) to[L,l_=\( L_{2b} \)] (2,0);
				\draw (-2,4) to[short,-] (-2,2);
				\draw (-2,2) 
				to[barrier,l_=\( \varphi_{1} \)] (-2,0);
				\draw (-2,2) 
				to[short,i_=\( I_{1} \)] (-2,1);
				\draw (2,2) 
				to[barrier,l^=\( \varphi_{2} \)] (2,0);
				\draw (2,2) 
				to[short,i^=\( I_{2} \)] (2,1);		
				\draw (2,4) to[short,-*] (2,2);		
				% \draw (0,4) to[L, ( L_{2a} \)] (-2,0);
				
				% Output lead
				\draw (0,-1) to[short,o-*] (0,0);
			\end{circuitikz}
			\caption{}
			\label{fig:biSQUID-diagram}
		\end{subfigure}
		\begin{subfigure}{0.49\textwidth}
			\begin{circuitikz}
				\draw (0,6) to [barrier] (0,0);
				\draw (-1.5,4.5) to[short,-*] (0,4.5) to (1.5,4.5);
				\draw (-1.5,4.5) to[R, l_=\( R \)] (-1.5,1.5);
				\draw (1.5,4.5) to[C, l=\( C \)] (1.5,1.5);
				\draw (-1.5,1.5) to[short,-*] (0,1.5) to (1.5,1.5);
			\end{circuitikz}
			\caption{}
			\label{fig:RCSJ}
		\end{subfigure}
		
		\caption{(a) Circuit representation of a bi-SQUID. The device is biased with an input current of \( I_b \). The (partial) self-inductances of the device are \( L_{1a},L_{1b},L_{2a},L_{2b},L_3 \) and the Josephson phases are \( \varphi_k \), \( k=1,2,3 \). (b) The RCSJ model for a Josephson junction, in which a physical Josephson junction is approximated by an equivalent circuit consisting of a resistor, capacitor and dissipation-less junction connected in parallel.}
		\label{fig:circuit-diagram}
	\end{figure*}}
	
	\subsection{Numerical Model}
	\label{sec:methodsB}
	The bi-SQUID model used in this work is based upon the lumped circuit element approximation, where the resistively and capacitively shunted junction (RCSJ) model is used to describe the Josephson junctions as circuit elements~\cite{Susan-PhD,Longhini2012}. One of the novelty of this model is that we are focusing on the simulation of a bi-SQUID with a model that account for change in critical current. The corresponding circuit diagrammatic representation of the bi-SQUID is shown in~\Cref{fig:biSQUID-diagram}.\\
	
	In this model, a Josephson junction is described by three circuit components connected in parallel: a shunt resistor, a capacitor and an element representing an `ideal', dissipation-less junction~\cite{Tinkham, Chesca2004}, as shown in the diagram in~\Cref{fig:RCSJ}. By an `ideal' junction, we refer to a circuit element in which the current \( I_j \) flowing through it is governed by a current-phase relation, a function of the gauge-invariant phase difference or Josephson phase~\( \varphi \) across the junction. The sinusoidal current-phase relation as below is imposed for all Josephson junctions in the model, which is valid for the long SNS-type junctions of the considered device~\cite{DeSimoni2022}
	
	\begin{equation}
		I_j(\varphi)=I_{c}\sin(\varphi)\cma
	\end{equation}
	where \( I_c \) is the critical current of the junction.
	
	The total current \( I_k \) flowing through the~\( k\textsuperscript{th} \) junction in the device (\( k=1,2,3 \)) is therefore given by the sum of the currents flowing through each parallel component
	
	\begin{equation}
		I_k(\varphi_k)=I_{c_k}\sin(\varphi_k)+\frac{V_k}{R_k}+C_k\dv{V_k}{t}\cma
	\end{equation}
	where \( V_k \) is the voltage across the junction, \( R_k \) is the shunt resistance and \( C_k \) is the junction capacitance.
	
	We rewrite the voltages in terms of the corresponding Josephson phases using the Josephson relation~\cite{Tinkham}
	\begin{equation}
		V_k=\frac{\hbar}{2e}\dv{\varphi_k}{t}\cma
	\end{equation}
	which results in
	\begin{equation}
		\label{eq:J-current} I_k(\varphi_k)=I_{c_k}\sin(\varphi_k)+\frac{\hbar}{2eR_k}\dv{\varphi_k}{t}+\frac{\hbar C_k}{2e}\dv[2]{\varphi_k}{t}\fstop
	\end{equation}
	
	A number of characteristic parameters are defined for the device so that~\cref{eq:J-current} for the Josephson current can be non-dimensionalized. All current terms are expressed in units of \( \ev{I_c}=\frac{1}{2}\qty(I_{c_1}+I_{c_2}) \), the mean critical current of the parallel junctions, labelled 1 and 2 as shown in~\Cref{fig:biSQUID-diagram}. Explicitly, in this work, each current \( I \) and critical current \( I_c \) in the system, are denoted by \( \iota \) and \( \iota_c \) to be the corresponding dimensionless current parameters, respectively. The junction resistances and capacitances receive a similar treatment; \( \ev{R} \) and \( \ev{C} \) represent the mean values of the resistance and capacitance of junctions 1 and 2, and this model introduce the dimensionless parameters \( r_k=\flatfrac{R_k}{\ev{R}} \) and \( c_k=\flatfrac{C_k}{\ev{C}} \), which represent the deviation of the respective parameter from the mean value.\\
	
	Following all of this, it is also natural to rescale time in terms of a characteristic time scale. Indeed, the non-dimensionalized time parameter~\( \tau=\omega t \) is also introduced, and defined with~\( \omega=\sqrt{\frac{2e\ev{I_c}}{\hbar\ev{C}}} \) is the characteristic frequency of the SQUID.\\
	
	The dimensionless form of~\cref{eq:J-current} thus reads
	
	\begin{equation}
		\label{eq:J-current-norm}
		\iota_k(\varphi_k)=\iota_{c_k}\sin(\varphi_k)+\frac{1}{Qr_k}\dv{\varphi_k}{\tau}+c_k\dv[2]{\varphi_k}{\tau}\cma
	\end{equation}
	with the introduction of~\( Q=\omega\ev{R}\ev{C} \) as the quality factor of the SQUID.\\

	In the circuit model, the dynamical behaviour of the bi-SQUID when it is subject to an external magnetic field is governed by the time evolution of the Josephson phases~\( \varphi_k \). If the time dependent behaviour of the~\( \varphi_k \) is known, it is then possible to compute the voltage across the device, which is the main observable of interest for this system under this mode of device operation. This will enable the modelling of the various characteristic curves of the device: namely, the voltage-current (V-I) and voltage-flux (V-B) curves.\\
	
	The Josephson phase dynamics of the bi-SQUID is governed by a system of 3 coupled second-order ordinary differential equations, which are obtained by combining~\cref{eq:J-current-norm} with current conservation equations obtained via the application of Kirchhoff's current laws in a circuit network analysis of the bi-SQUID circuit from~\Cref{fig:biSQUID-diagram}, and a pair of additional constraint equations that relate the Josephson phases to the magnetic flux threading the loops in the device (one for each loop), called the~\emph{flux quantisation conditions}. The flux quantisation conditions for the bi-SQUID read
	
	\begin{align}
		\varphi_1-\varphi_2&= 2\pi\frac{\Phi^{\text{total}}}{\Phi_0}\quad (\text{mod }2\pi)\\
		\varphi_1-\varphi_2+\varphi_3&= 2\pi\frac{\Phi^{\text{lower}}}{\Phi_0}\quad (\text{mod }2\pi)\cma
	\end{align}
	where \( \Phi^{\text{total}} \) is the total magnetic flux contained in the device, and \( \Phi^{\text{lower}} \) is the flux through just the lower loop. The flux terms can be decomposed into contributions from the external magnetic field, as well as self-inductive couplings due to current circulating the device
	
	\begin{align}
		\Phi\textsuperscript{total}&= \Phi^{\text{total}}_{\text{ext}} -L_{1a}I_L+L_{1b}I_R+L_{2b}I_2-L_{2a}I_1\\
		\Phi\textsuperscript{lower}&= \Phi^{\text{lower}}_{\text{ext}}-L_3I_3 +L_{2b}I_2 -L_{2a}I_1\fstop
	\end{align}
	
	These relations allow for all of the currents \( I_L, I_R, I_1, I_2, I_3 \) to be eliminated, resulting in a system of governing equations in terms of the Josephson phases as the only dynamical variables, this is shown in detail within ~\cref{Appendix}. This system of differential equations is solved numerically using typical \textsc{Matlab} ODE solver routines to obtain the time evolution of the \( \varphi_k \)~\cite{Susan-PhD}. This is achieved through the use of the inbuilt \textsc{ode45} solver to integrate a system of differential equations from an initial time $t_0$ to a final time $t_f$ with a certain tolerance error. For example, typical value of the relative tolerance is $10^{-6}$ and can can be refined to $10^{-8}$ in some scenarios.\\
	
	The average voltage response that would be measured across the device is then computed via a time-average of the Josephson phases \( \varphi_1 \) and \( \varphi_2 \)~\cite{Susan-PhD}
	%\textbf{I suspect that some part of this must be removed, the details about ~\( \tau_f \) must be removed}
	\begin{equation}
		\label{eq:avV} \ev{V}=\frac{\hbar\omega}{2e}\lim_{\tau_f\to\infty}\qty(\frac{1}{\tau_f}\int_0^{\tau_f}\frac{1}{2}\qty(\dv{\varphi_1}{\tau}+\dv{\varphi_2}{\tau})\dd{\tau})\fstop
	\end{equation}
	
	%The time average is evaluated by computing the mean of the solution arrays for~\( \dv{\varphi_1}{\tau} \) and~\( \dv{\varphi_2}{\tau} \) over a truncated time interval~\( [0, \tau_f] \) for a sufficiently large~\( \tau_f \).\\
	
	The output of the model described in this paper therefore depends on the specification of the following constants: the junction critical currents \( I_{c_k} \), resistance~\( R_k \) and capacitance~\( C_k \), the input bias current~\( I_b \), the (partial) self-inductances~\( L_{1a}, L_{1b}, L_{2a}, L_{2b}, L_3 \), the external flux~\( \Phi^{\text{total}}_{\text{ext}} \), and the loop areas. Most of these parameters can be obtained via direct measurement. The inductances are computed using a commercial 3D superconducting circuit parameter extraction software called \emph{InductEx}~\cite{Fourie2011}. This allows for the use of real parameters within modelling in order to draw acceptable conclusions.\\

	As a time dependent input bias current is used in the measurements discussed in this work, the following adjustments are made to the method in order to compute the voltage response. A particular sinusoidal input of frequency \( \nu \) and amplitude \( A \) are used for the input bias as described below:
	\begin{equation}
		I_b(t)=A\sin(2\pi\nu t)\fstop
	\end{equation}
	the sinusoid above is uniformly sampled over a single period to obtain an array of \( I_b \) values. \( \ev{V} \) is then computed by using~\cref{eq:avV} at each sample point. These computations are performed in time-order, where the final state of the solution for one calculation is used as the initial condition for the following computation. It can be assumed that the system has adequate time to settle into its stable long-term behaviour between each sample point, provided that the frequency of the input bias current signal is negligibly small in comparison to the characteristic frequency \( \omega \) of the SQUID (\( \nu\ll\omega \)). Then by averaging over all sample points, it is possible to obtain the time averaged voltage for the time dependent input. This method allows for the V-B behaviour of the device to be sensitive to the hysteretic nature of its V-I characteristic; the device critical current will be different for the forward and backward sweeps of the bias current. The approach described above is somehow in line with a rich list of examples presented in the literature that aims at a more realistic description of dynamical effects in Josephson systems; e.g.~\cite{EmmaPRB2021} \\
	
	As a next step used to model the novel effects at the centre of this study, to consider the electrostatic gating effect in the model, the current-phase relation of each gated Josephson junction is modified by promoting the critical current to be a function of the gate voltage \( V_g \)
	\begin{equation}
		I_k=I_c(V_g)\sin(\varphi_k)\fstop
	\end{equation}
	
	The \( I_c \) vs.\ \( V_g \) curve is captured by two additional parameters: a threshold voltage~\( V_{\text{thres}} \) when critical current suppression is observed for~\( \abs{V_g}\geq V_{\text{thres}} \), and a second (`saturation') threshold \( V_{\text{sat}} \) after which the critical current is completely suppressed. It is assumed that the curve obeys a linear relationship in the region~\( V_{\text{thres}}\leq \abs{V_g}\leq V_{\text{sat}} \), and that the gating effect is bipolar, i.e.\ symmetric under exchange of the sign of the voltage \( V_g \), which is consistent with experimental observations~\cite{DeSimoni2018, DeSimoni2021b, Rocci2020, DeSimoni2019}. Hence, this behaviour may be captured in the following semi-empirical form for the gate voltage dependence of the critical current
	
	\begin{equation}
		I_c(V_g)=I_c(0)\qty[1-\frac{\theta(\Delta V_t)}{V_{\text{sat}}-V_{\text{thres}}}\Delta V_t](1-\theta(\Delta V_s))\cma
	\end{equation}
	where \( \theta \) is the Heaviside step function,~\( \Delta V_t=\abs{V_g}-V_{\text{thres}} \) and~\( \Delta V_s=\abs{V_g}-V_{\text{sat}} \). %This proposed relationship is based upon the assumption that the gating effect is bipolar, i.e.\ symmetric under exchange of the sign of the voltage \( V_g \), which is consistent with experimental observations~\cite{DeSimoni2018, DeSimoni2021b, Rocci2020, DeSimoni2019}. 
	The values of~\( V_{\text{thres}} \) and~\( V_{\text{sat}} \) will be obtained from fits to experimental data. The maximum, unsuppressed value for the critical current \( I_c(0) \) will simply be the value of the critical current of the ungated Josephson junction. With the above a bi-SQUID with gate-controlled Josephson junctions may be modelled without the presence of a microscopic theory, that has not yet been developed, to describe fully how gating Josephson junctions will affect a bi-SQUID.
	
	\section{Results}
	\label{sec:results}
	Crude estimates for the critical current and the resistance were obtained from the experimental V-I curve using the superconducting-dissipative transition point and the slope in the dissipative regime, respectively. Assuming identical junctions, we obtain a junction normal state resistance of \( \sim 77.76\,\Omega \) and a critical current of \( \sim 13.69\,\mu\text{A}\). The ratio of the third junction critical current \( I_{c_3} \) to the average un-gated critical current \( \ev{I_c} \) is estimated to be \( 0.5 \), which is consistent with values obtained for a previous similar device\cite{DeSimoni2022}. Since the junctions are operating in the overdamped regime, we set the junction capacitance to \( 0.5 \) fF, which accordingly sets the Stewart-McCumber parameter \( \beta_c=\frac{2\pi}{\Phi_0}I_cR^2C\sim 0.1\ll 1 \) as required\cite{Chesca2004}.\\
	
	Values for the self-inductances were extracted using InductEx using information about the geometry of the device and its material and layer composition, from which we obtain \( L_{1a}=8.47 \) pH, \( L_{1b}=8.47 \) pH, \( L_{2a}=1.84 \) pH, \( L_{2b}=1.82 \) pH and \( L_3=1.57 \) pH.\\ 
	
\begin{figure*}
	\includegraphics[width=\linewidth]{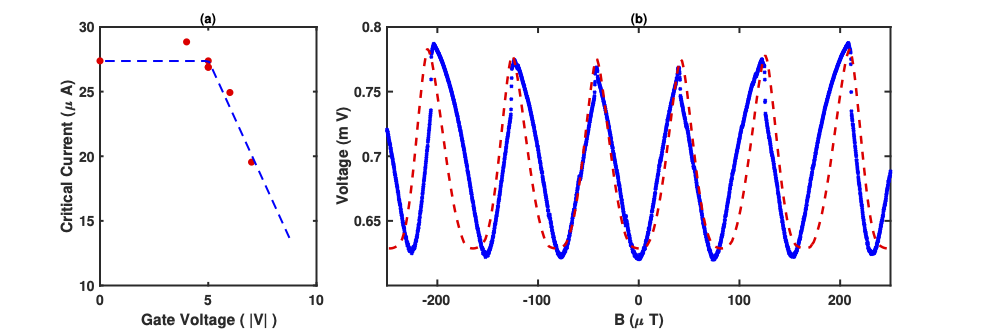}
	\caption{(a) Switching current of a gated SNS bi-SQUID at 30 mK vs.\ the applied gate voltage. Experimental data is marked with red dots. The fitted curve to our theory is drawn with a dashed line. (b) Simulated V-B curve of the SNS bi-SQUID biased with a current of \( I_b(t)=23\sqrt{2}\sin(2\pi\times 17\text{ Hz}\times t)\,\mu \)A for \( V_g= 0 \) V (red dashed line), alongside corresponding experimental data (blue crosses).}
		\label{fig:sim}
	\end{figure*}

	From the measured V-I curves of the device characterised in the absence of magnetic fields, it is possible to extract the switching current for each applied gate voltage in the dataset. This approach provides the experimental values for the device critical current vs.\ gate voltage. The performing of a nonlinear least squares fit of the data (in the region where critical current suppression is observed) is used to allow the extraction of values for~\( V_{\text{thres}} \) and~\( V_{\text{sat}} \). A first-order polynomial for the fit function is used for this task and values \( V_{\text{thres}}=5.02 \) V and \( V_{\text{sat}}=8.75 \) V are extracted from this procedure. The modelled relationship between the device critical current vs.\ \( V_g \) alongside the corresponding experimental data is then plotted in \Cref{fig:sim}.\\
	
\begin{figure*}
		\begin{subfigure}{\textwidth}
			\includegraphics[width=\linewidth]{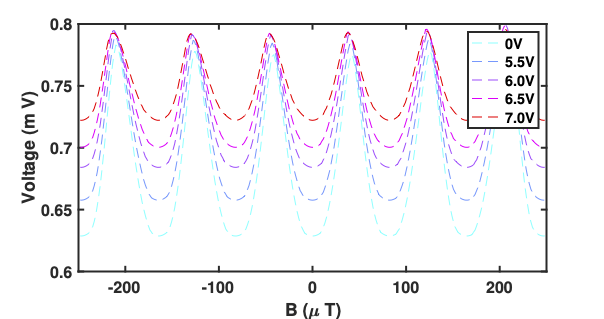}
		\end{subfigure}
		\begin{subfigure}{\textwidth}
			\includegraphics[width=\linewidth]{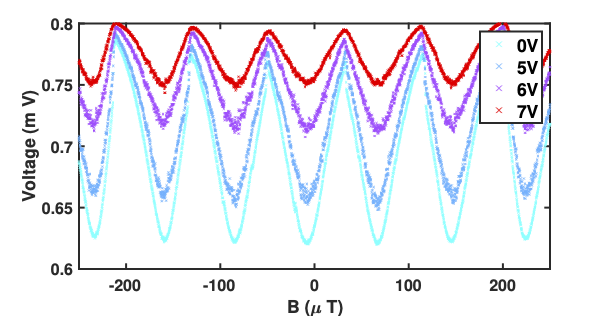}
		\end{subfigure}
		\caption{V-B curves for the SNS bi-SQUID biased with a current of \( I_b(t)=23\sqrt{2}\sin(2\pi\times 17\text{ Hz}\times t)\,\mu \)A for various gate voltage values ranging from 0 to 7V. Top: simulation results. Bottom: experimental data. The exploration of the linearity of the above curves has been excluded from this paper although an initial increase in linearity was observed for small gate voltage until a threshold voltage.}
		\label{fig:main}
	\end{figure*}

	With the required device parameters now completely specified, it is possible to proceed with the modelling of the V-B curves when the device is biased with a \( 17 \)~Hz sinusoidal input current with an RMS magnitude of 23~\( \mu \)A (which corresponds to an amplitude of~\( 23\sqrt{2} \mu\text{A}\sim 32 \mu\)A). A flux range equal to~\( \pm 3\Phi_0 \), corresponding to the range of data experimentally available, is investigated. A theoretical plot of the V-B curve where the junctions are un-gated (\( V_g=0 \) V) is shown in~\Cref{fig:sim}, and a similar theoretical plot of the V-B curves for the gate voltages 0, 5.5, 6.0, 6.5 and 7.0 V is shown in~\Cref{fig:main}. The corresponding experimental data measured at 30 mK are shown alongside for ease of comparison. A constant horizontal offset of \(7.6\,\mu \)T is applied to the experimental results to recenter the central anti-peak on \( B=0 \), and a constant voltage offset of~\( 0.045 \) mV is applied to the simulation results to obtain even better agreement with the experimental results. The validity of such modifications will be discussed further in \Cref{sec:discussion}.
	
	\section{Discussion}
	\label{sec:discussion}
	
	From the plot of the un-gated V-B curves in~\Cref{fig:sim}, it is possible to observe that the modified RCSJ model introduced in this work does captures the basic behaviour of the SNS bi-SQUID to a good extent, successfully reproducing the overall shape, amplitude and flux-periodicity of the characteristic curve. The minor periodicity mismatch can be attributed to uncertainties induced through fabrication leading to non-exact parameters used for modelling (size of the loop, total inductances, etc.).\\
	
	%The constant horizontal and vertical offsets that were applied serve to take into account systematic uncertainties that are inevitably introduced during the fabrication of the device and the measurement process. The re-centring of the experimental V-B curve removes any stray flux sources not due to the superconducting electromagnet, ensuring the correct calibration of the horizontal scale of the V-B curve. The requirement of a voltage offset to better capture the data may be physically attributed to a constant uncertainty in the voltage measurement, i.e.\ due to instrumentation error or to some imperfections in the fabrication that may bring some errors in the extraction of the physical parameters.\\
	
	Our proposed relationship between the switching current of the device and the applied gate voltage displays reasonable agreement with the experimental data, modulo a single outlier, as seen in~\Cref{fig:sim}. We therefore rely on the fact that its behaviour is consistent with previous experimental observations~\cite{DeSimoni2018,DeSimoni2019}, and so it is an appropriate model for our device under consideration. \\
	
	%%\textcolor{red}{I removed the line from here that I think Giogrio was referring to and feel like its now missing a sentence but not sure what to say.}\\
	
	The V-B curves in~\Cref{fig:main} modelled with a gate voltage-dependent critical current term possess many of the essential features of the gated V-B data. As with the un-gated data, the flux-periodicity of the simulated curves is correct. The simulations also correctly capture the overall trend of the decreasing response amplitude with increasing gate voltage by virtue of the decrease of the switching current. It should be noted that the modelling was performed without the addition of stochastic noise and fluctuations, which may manifest in the experiment from electrical and thermal sources. This accounts for one of the main visual differences between the model and experiment, and may also be a factor in causing the slight discrepancy between the response amplitudes of the curves, particularly for the higher \( V_g \) values. Aside from these differences, that go beyond the scope of the focus of this paper, the model displays a reasonable level of qualitative agreement.\\
	
	The main result proved with these data is that the only major influence of the application of the gate voltage on the device is a reduction in the junction critical current opens the possibility of exploiting the gating effect to improve the operation of SQUIDs in sensing applications. SQUID sensors are carefully engineered, requiring precise control of its physical parameters in order to optimise its V-B transfer function in terms of linearity and sensitivity. While there exist advanced specialised fabrication techniques in order to minimise the random variation of its physical parameters from their design values, the total elimination of such flu ctuations is unfeasible.\\
	
	However, it is possible to tune the junction critical currents~\emph{after} they have been fabricated via judicious application of electrostatic gating; the gate voltage can be treated as a `dial' to set the junction critical currents to their desired values, or even varied in real time during operation until the most optimal output is obtained. Since the applied voltage does not appear to result in any other major side effects on the device behaviour, this method may provide a less strict tolerance for parameter control during fabrication~\cite{patent}.

	\section{Conclusions}
	In this work it is therefore demonstrated that the behaviour of an SNS bi-SQUID can be adequately captured by a modified RCSJ circuit model, noting that our results are obtained by using parameters very close to the real ones. Our work demonstrates that the gating effect can be successfully accounted for in a minimal extension in the circuit theory framework by modifying the junction critical current to be a decreasing function of the magnitude of the gate voltage.\\
	
	These results suggests that the gating effect does not influence any other aspects of the device to a significant extent. The effect can therefore be exploited to individually tune the junction critical currents in the device so that they can be freely set to their optimal values post-fabrication, thus avoiding the difficulties associated with random variation of physical parameters during fabrication~\cite{patent}. Although this was only explicitly verified for a bi-SQUID in this work, it is likely to also be applicable to any superconducting device constructed with SNS junctions. By making this simplification an initial model could be used in order to determine the affects on the system of gated Josephson junction. Because sophisticated microscopic theories are often difficult to deploy in an efficient way for such complicated modelling tasks~\cite{Tinkham}, the results contained in this paper may suggest pathways for a more simplistic use of these complicated microscopic models and the development of a more detailed and accurate model.

	\bibliography{main_edit}
	
	\appendix*
	\section{Bi-SQUID equations final form}
	\label{Appendix}
	The final form of the bi-SQUID equations may be found using the following equations from the main text, firstly the flux quantisation conditions
	\begin{align}
		\varphi_1-\varphi_2&= 2\pi\frac{\Phi^{\text{total}}}{\Phi_0}\quad (\text{mod }2\pi)\label{A1}\\
		\varphi_1-\varphi_2+\varphi_3&= 2\pi\frac{\Phi^{\text{lower}}}{\Phi_0}\quad (\text{mod }2\pi)\cma\label{A2}
	\end{align}
	and the relation between external flux and flux as a result of self inductance
	\begin{align}
		\Phi\textsuperscript{total}&= \Phi^{\text{total}}_{\text{ext}} -L_{1a}I_L+L_{1b}I_R+L_{2b}I_2-L_{2a}I_1\label{A3}\\
		\Phi\textsuperscript{lower}&= \Phi^{\text{lower}}_{\text{ext}}-L_3I_3 +L_{2b}I_2 -L_{2a}I_1\fstop\label{A4}
	\end{align}
	The current conservation laws are also required in order to construct the governing equations which are given by 
	\begin{align}
		I_{b} &= I_{L}+I_{R}\label{A5}\\
		I_{L}+I_{3} &= I_{1}\label{A6}\\
		I_{R} &= I_{3}+I_{2}\label{A7}\cma
	\end{align}
	together with the RCSJ equation in dimensionless form
	\begin{equation}
		\iota_k(\varphi_k)=\iota_{c_k}\sin(\varphi_k)+\frac{1}{Qr_k}\dv{\varphi_k}{\tau}+c_k\dv[2]{\varphi_k}{\tau}\fstop\label{A8}
	\end{equation}\\
	
	To obtain the first of the two governing equations consider the flux through the total loop given by \cref{A1} and substitute \cref{A3}
	\begin{equation}
			\varphi_1 + L_{1a}\iota_L + L_{2a}\iota_1 = \Phi^{\text{total}}_{\text{ext}} + \varphi_2 + L_{1b}\iota_R + L_{2b}\iota_2\cma
	\end{equation}
	where the factor of $\flatfrac{2\pi}(\Phi_{0})$ has been absorbed into $\Phi^{\text{total}}_{\text{ext}}$ and all of the $L$ such that the equation is dimensionless. And now applying the first two current conservation~\cref{A5} and~\cref{A6} to eliminate $\iota_{R}$
	\begin{align}
		(L_{1a}+L_{1b}+L_{2a})\iota_{1} = \Phi^{\text{total}}_{\text{ext}} + \varphi_2 - \varphi_1 + L_{1b}\iota_b + L_{2b}\iota_2\nonumber \\
		+ (L_{1a}+L_{1b})\iota_{3}\cma\label{A10}
	\end{align}
	the equation is now written in terms only of currents through Josephson junctions and the bias current. Using~\cref{A8} the Josephson currents can now be written in terms of the Josephson phases giving the first of the bi-SQUID final form equations. The second final form equation is obtained similarly by substituting~\cref{A3} into~\cref{A1} once again, this time applying the current conservation~\cref{A5} and~\cref{A7} to eliminate $\iota_{L}$
	\begin{align}
		-(L_{1a}+L_{1b}+L_{2b})\iota_{2} = \Phi^{\text{total}}_{\text{ext}} + \varphi_2 - \varphi_1 - L_{1a}I_{b} - L_{2a}\iota_1 \nonumber \\
		+ (L_{1a}+L_{1b})\iota_{3}\fstop \label{A11}
	\end{align}
	As before~\cref{A8} can now be used to eliminate currents resulting in the second bi-SQUID final form equation.
	To obtain the third and final bi-SQUID equation the lower loop only is now considered by substituting~\cref{A2} into~\cref{A4}
	\begin{equation}
		\varphi_1-\varphi_2+\varphi_3 = \Phi^{\text{lower}}_{\text{ext}} -L_{3}\iota_3+L_{2b}\iota_2-L_{2a}\iota_1\fstop\label{A12}
	\end{equation}\\
	
	As the only currents present in~\cref{A10,A11,A12} are those through Josephson junctions,~\cref{A8} can be applied to obtain the final full form bi-SQUID equation. Using~\cref{A8} to rewrite the above in terms of the Josephson phases as the only dynamical variable gives the resulting three governing equations 
	\begin{align}
		(L_{1a}+L_{1b}+L_{2a})\left(\iota_{c_1}\sin(\varphi_1)+\frac{1}{Qr_1}\dot{\varphi}_1+c_1\ddot{\varphi}_1\right) \nonumber \\
		- L_{2b}\left(\iota_{c_2}\sin(\varphi_2)+\frac{1}{Qr_2}\dot{\varphi}_2+c_2\ddot{\varphi}_2\right) \nonumber \\
		- (L_{1a}+L_{1b})\left(\iota_{c_3}\sin(\varphi_3)+\frac{1}{Qr_3}\dot{\varphi}_3+c_3\ddot{\varphi}_3\right) \nonumber \\
		= \Phi^{\text{total}}_{\text{ext}} + \varphi_2 - \varphi_1 + L_{1b}\iota_b\fstop \\
		 L_{2a}\left(\iota_{c_1}\sin(\varphi_1)+\frac{1}{Qr_1}\dot{\varphi}_1+c_1\ddot{\varphi}_1\right) \nonumber \\
		-(L_{1a}+L_{1b}+L_{2a})\left(\iota_{c_2}\sin(\varphi_2)+\frac{1}{Qr_2}\dot{\varphi}_2+c_2\ddot{\varphi}_2\right) \nonumber \\ 
		- (L_{1a}+L_{1b})\left(\iota_{c_3}\sin(\varphi_3)+\frac{1}{Qr_3}\dot{\varphi}_3+c_3\ddot{\varphi}_3\right) \nonumber \\
		= \Phi^{\text{total}}_{\text{ext}} + \varphi_2 - \varphi_1 - L_{1a}\iota_b\fstop \\
		 L_{2a}\left(\iota_{c_1}\sin(\varphi_1)+\frac{1}{Qr_1}\dot{\varphi}_1+c_1\ddot{\varphi}_1\right) \nonumber \\
		-L_{2b}\left(\iota_{c_2}\sin(\varphi_2)+\frac{1}{Qr_2}\dot{\varphi}_2+c_2\ddot{\varphi}_2\right) \nonumber \\ 
		+L_3\left(\iota_{c_3}\sin(\varphi_3)+\frac{1}{Qr_3}\dot{\varphi}_3+c_3\ddot{\varphi}_3\right) \nonumber \\
		= \Phi^{\text{total}}_{\text{ext}} + \varphi_2 - \varphi_1 - \varphi_3\fstop
	\end{align}
		
\end{document}